\pdfoutput=1
\documentclass[letterpaper]{article}
\usepackage[preprint]{aaai2027}
\usepackage[hyphens]{url}
\usepackage{graphicx}
\usepackage{natbib}
\usepackage{caption}
\usepackage{booktabs}
\usepackage{multirow}
\usepackage{amsmath}
\usepackage{amssymb}
\usepackage[hidelinks,bookmarksnumbered]{hyperref}
\newcommand{\ModuleName}{VividAudio}

\newcommand{\AugName}{IntraEvtVar}

\newcommand{\EvalName}{VGGSound-Ref}
\newcommand{\VividBench}{VividBench}

\title{\ModuleName: Multi-Event Reference-Conditioned Video-to-Audio Generation}
\author{
  You Li\textsuperscript{1,2,3} \quad
  Dewei Zhou\textsuperscript{1,2} \quad
  Fan Ma\textsuperscript{1,2} \quad
  Dongliang He\textsuperscript{3} \quad
  Yi Yang\textsuperscript{1,2,\dag}
}
\affiliations{
  {\small \textsuperscript{1}The State Key Lab of Brain-Machine Intelligence, Zhejiang University}\\
  {\small \textsuperscript{2}ReLER, CCAl, Zhejiang University, China}\\
  {\small \textsuperscript{3}ByteDance, China \quad
  \textsuperscript{\dag}Corresponding author}\\
  {\ttfamily\small \{uli2000, zdw1999, mafan, yangyics\}@zju.edu.cn \quad hedlcc@126.com}
}

\hypersetup{
  pdftitle={VividAudio: Multi-Event Reference-Conditioned Video-to-Audio Generation},
  pdfauthor={You Li, Dewei Zhou, Fan Ma, Dongliang He, Yi Yang},
  pdfsubject={Reference-conditioned video-to-audio generation}
}

\makeatletter
\let\aaai@orig@maketitle\@maketitle
\def\@maketitle{%
  \aaai@orig@maketitle
  \noindent\begin{minipage}{\textwidth}
    \centering
    \includegraphics[width=0.7\textwidth]{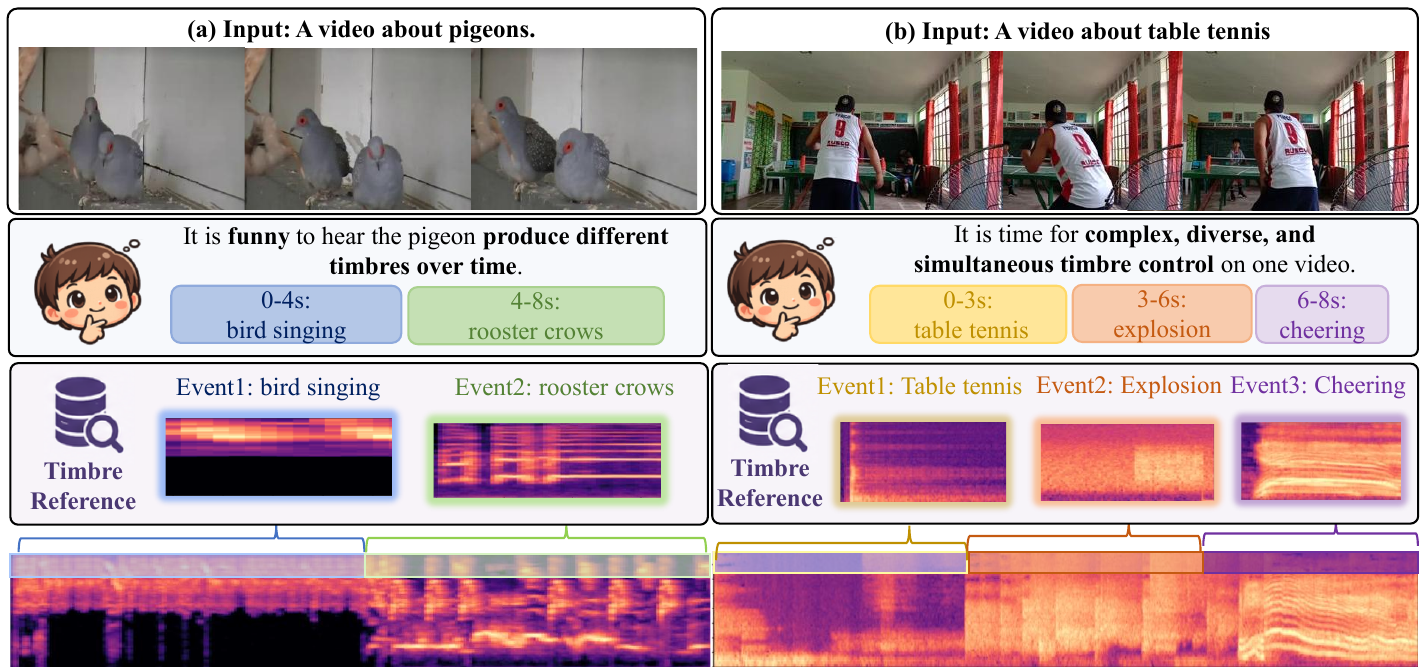}
    \captionof{figure}{\textbf{\ModuleName{} enables reference-conditioned Foley generation beyond semantic V2A.}
    Audio exemplars steer the acoustic identity of the same visual event, separate references control distinct events in multi-event scenes.}
    \label{fig:teaser}
  \end{minipage}
}
\makeatother

\begin{document}
\maketitle
\thispagestyle{plain}
\begin{abstract}
Existing video-to-audio (V2A) methods have achieved promising results in synthesizing high-quality audios. However, as V2A advances toward more flexible user-directed creation, existing methods remain notably limited in controlling the distinct acoustic identities of multiple events within a video. We therefore formulate \emph{multi-event reference-conditioned V2A} (Mr-V2A), where each event is assigned a specific audio reference and target time span. Mr-V2A presents three key challenges. First, event-level reference-target training pairs are scarce. We introduce \emph{Intra-Video Event Variation} (\AugName{}), which constructs non-oracle pairs through four variation modes of events mined within each video. Second, we organize the references into a \textbf{T}ime-\textbf{A}ligned \textbf{CO}mposite \textbf{Ref}erence (TACO-Ref) and inject it through local and global pathways: the local pathway retains time-resolved event cues, while an Audio Q-Former provides a clip-level acoustic-context condition. Third, jointly conditioning on multiple references can cause acoustic-attribute leakage across events. Localized reference attention and the TACO-Ref Router (TRR) explicitly constrain event-to-reference correspondence to reduce such interference. Experiments on \EvalName{} and \VividBench{} demonstrate stronger reference controllability than existing V2A and exemplar-based methods while maintaining competitive Foley generation quality.
\end{abstract}

\section{Introduction}
\label{sec:intro}
Recent advances in generative models~\cite{wan,cogvideox,flux,stablediffusion} have enabled realistic image and video synthesis, while video-to-audio (V2A) models~\cite{mmaudio,hunyuanfoley,kling,dreamfoley,thinksound} have made substantial progress in synthesizing Foley sounds that are semantically and temporally consistent with visual content.

As V2A advances toward more flexible user-directed creation, producing a plausible foley is no longer sufficient: creators may also want to specify \emph{what} sound events should occur, \emph{when} they should occur, and \emph{how} each event should sound. Existing controllable V2A methods address parts of this goal. Video-Foley~\cite{videofoley} and FoleyDirector~\cite{FoleyDirector} improve semantic and temporal controllability through event conditions, structured scripts, and temporal cues, while AC-Foley~\cite{acfoley} introduces reference audio for acoustic controllability in single-event Foley generation. However, real-world videos often contain multiple events that require distinct acoustic identities within one synchronized soundtrack. Existing methods remain notably limited in this multi-event setting.

\paragraph{Mr-V2A.}
We therefore formulate \textbf{M}ulti-event \textbf{r}eference-conditioned V2A (Mr-V2A). Given a video and a set of audio references paired with target time spans, Mr-V2A generates a synchronized soundtrack in which each event follows the acoustic identity of its designated reference, as illustrated in Fig.~\ref{fig:teaser}. This setting introduces three challenges. \textbf{1) Data limitation.} Existing datasets rarely provide $(\text{video}, \text{per-event reference}, \text{target audio})$ triplets, while using the target audio itself as the reference creates an oracle shortcut. \textbf{2) Multi-timbre representation and injection.} How can the multi-timbre be effectively encoded and integrated into the generation model without compromising video grounding?
\textbf{3) Timbre disentanglement.} Joint conditioning on multiple references can leak acoustic attributes across event, causing an event to follow the wrong reference.

\paragraph{Our Method.}
To address these challenges, we propose \textbf{\ModuleName} with following components. \textit{\textbf{i)} Intra-Video Event Variation (\AugName)}.  mines events recorded within the same video and constructs non-oracle reference-target pairs through group replacement, speed variation, mixing, and acoustic degradation. \textit{\textbf{ii)} Dual-Path TACO-Ref Injection.} we place the event references at their target positions to form a \textbf{T}ime-\textbf{A}ligned \textbf{CO}mposite \textbf{Ref}erence (TACO-Ref). Its time-resolved and Audio-Q-Former-aggregated features are injected through local and global pathways, respectively, to model event-specific acoustic identity and shared acoustic context.  \textit{\textbf{iii)} Attribute Disentanglement.} Localized Ref-Attn limits irrelevant temporal interactions, while the TACO-Ref Router (TRR) explicitly routes each event-specific response to its time span, reducing cross-event leakage.

\paragraph{Benchmarks.}
We establish two complementary benchmarks. \EvalName{} is an event-annotated held-out VGGSound split that evaluates reference-conditioned Foley generation under the training distribution. \VividBench{} evaluates a more realistic retrieval setting: we manually annotate event categories and time spans in $50$ videos, retrieve external references from Freesound, and measure whether each generated event follows its designated reference.

\paragraph{Evaluation.}
Experiments on \EvalName{} and \VividBench{} show that \ModuleName{} substantially improves reference controllability over both plain V2A and single-embedding exemplar baselines. On \EvalName{}, it raises BEATs-Cos/CLAP-Cos to $0.695/0.629$ while maintaining competitive quality under FAD$_{\text{VGG}}$/$\mathrm{KL}_{\text{PANNs}}$ ($0.839/1.62$). On \VividBench{}, the event-level gains become more pronounced as the number of acoustic identities increases.

\noindent Our main contributions are summarized as follows.

\begin{itemize}
    \item We formulate Mr-V2A, extending reference-conditioned Foley generation from a single exemplar to event-specific acoustic control over multiple events in one synchronized soundtrack.
    \item We propose \ModuleName{}, which addresses scarce supervision, multi-scale reference conditioning, and cross-event interference through \AugName{}, dual-path TACO-Ref injection, localized reference attention, and TRR.
    \item We develop a data construction pipeline together with \EvalName{} and \VividBench{}, and demonstrate improved, balanced multi-event reference controllability with competitive generation quality.
\end{itemize}

\section{Related Work}
\label{sec:related}

\subsection{Video-to-Audio}
Video-to-audio (V2A) extends audio generation with visual semantics and temporal structure. FoleyCrafter~\cite{foleycrafter}, Diff-Foley~\cite{difffoley}, Frieren~\cite{frieren}, VTA-LDM~\cite{vta-ldm}, and V-AURA~\cite{V-AURA} generate synchronized sound through multimodal diffusion or flow-based fusion. MultiFoley~\cite{multifoley} and MMAudio~\cite{mmaudio} combine text, audio, and video conditions for flexible synthesis, while ThinkSound~\cite{thinksound}, HunyuanVideo-Foley~\cite{hunyuanfoley}, and Kling-Foley~\cite{kling} scale V2A toward reasoning or unified multimodal generation. However, most systems still rely on global text and visual evidence, which are effective for semantic alignment but weak at specifying timbre, material, recording condition, and scene ambience.

\begin{figure*}[t]
    \centering
    \includegraphics[width=0.92\textwidth]{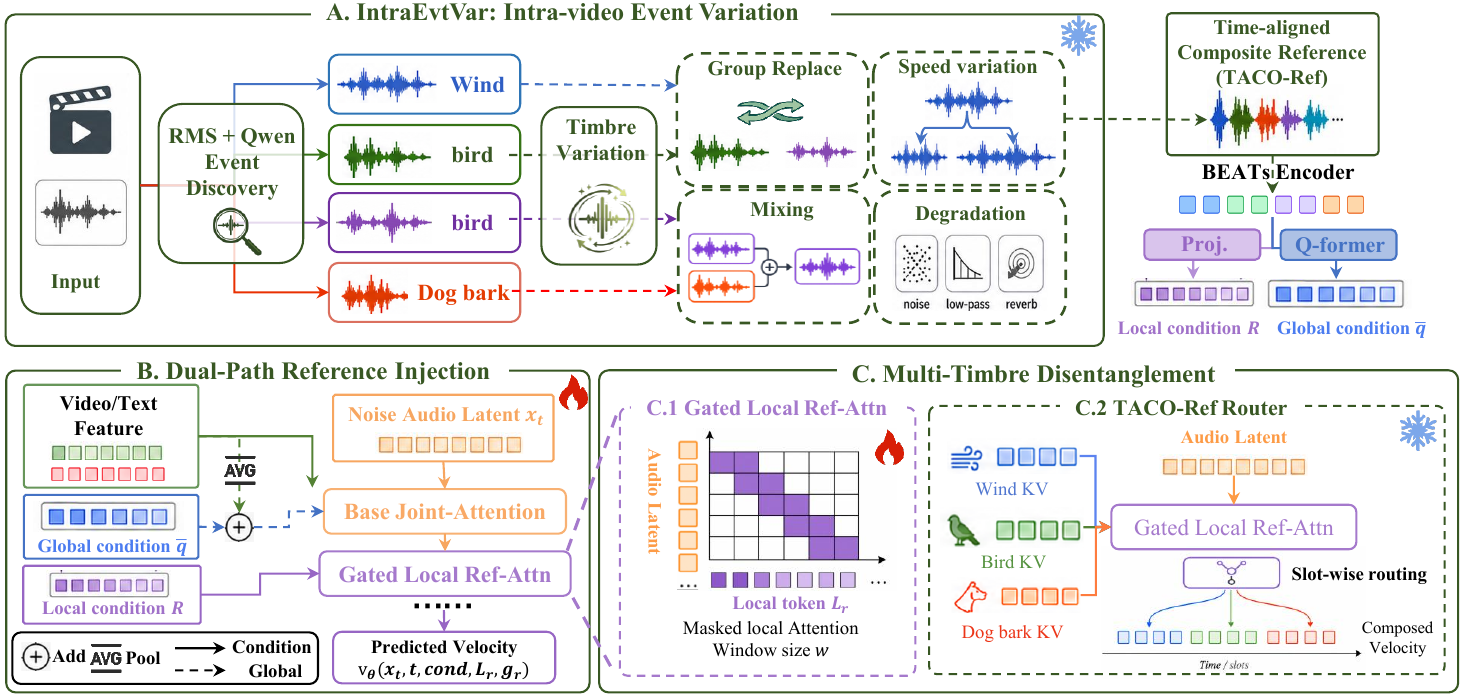}
    \caption{\textbf{Overview.}
    We first discover audio events within each video and explore four modes to construct reference-target pairs. The resulting TACO-Ref is injected into the pretrained V2A DiT through Dual-Pathways. Localized reference attention and the TACO-Ref Router (TRR) disentangle event-specific acoustic attributes.}
    \label{fig:method_overview}
\end{figure*}

DreamFoley~\cite{dreamfoley} further explores autoregressive V2A generation with vision-language models, jointly modeling video, audio, and text through discrete audio tokens.

\subsection{Controllable V2A}
Beyond unconditional or text-conditioned V2A, recent methods provide more direct control over Foley generation. Video-Foley~\cite{videofoley} and FoleyDirector~\cite{FoleyDirector} specify what sound events should occur and when they should be rendered, while MultiSoundGen~\cite{multisoundgen}, SelVA~\cite{SelVA}, and Foley-Flow~\cite{Foley-Flow} improve multi-event generation, selective source synthesis, and semantic-rhythmic synchronization. These methods strengthen control over \emph{what} to generate and \emph{when} to generate it, but do not explicitly control the acoustic identity of each event with reference audio.
Audio-conditioned V2A further studies \emph{how} a sound should sound. Tri-Ergon~\cite{Tri-Ergon} uses auditory prompts for global style and loudness control, FoleyGenEx~\cite{FoleyGenEx} introduces reference-audio injection in a unified multimodal framework, and AC-Foley~\cite{acfoley} uses audio exemplars for acoustic transfer. However, their audio condition is mainly used as a global style/acoustic cue, where timbre, event semantics, and temporal patterns may remain entangled. In contrast, Mr-V2A assigns distinct references to individual events and disentangles multiple acoustic identities through temporally localized reference routing.

\subsection{Related Visual and Media Tasks}
Fine-grained control is also studied in visual generation. MIGC~\cite{migc} and MIGC++~\cite{migc++} investigate multi-instance image generation, while RefineAnything~\cite{refineanything} studies region-specific image refinement with background preservation. Complementary uses of generative models include synthetic data construction for vision-language tasks in AnySynth~\cite{li2024anysynth} and imagined proxies for composed image retrieval in Imagine and Seek~\cite{ipcir}. In media production, RoughCut-Bench~\cite{roughcutbench} evaluates vision-language models on automatic video rough-cut and structured cut planning. These tasks provide broader context for controllable media creation; our focus is event-specific acoustic identity within a video-synchronized soundtrack.

\section{Method}
\label{sec:method}

\subsection{Definition and Overview}
\paragraph{Definition.}
Given a video $V$, an optional text prompt $c$, and a set of event-level reference conditions
$\mathcal{R}=\{(a_k^r,\tau_k)\}_{k=1}^{K}$, where $a_k^r$ is a reference audio clip and $\tau_k$ specifies the target time span in which its acoustic identity should be expressed, \ModuleName{} generates a waveform $\hat a$ that remains synchronized with the visual events while preserving the assigned reference timbre within each corresponding time span. We build on an MMAudio-style~\cite{mmaudio} multimodal DiT and train it with rectified flow matching in the latent space of a $44$\,kHz VAE:
\[
\mathcal{L}_{\text{FM}}(\theta) = \mathbb{E}_{\mathbf{x}_0, \mathbf{x}_1, t}\big[\lVert v_\theta(\mathbf{x}_t, t, \mathbf{c}) - (\mathbf{x}_1 - \mathbf{x}_0) \rVert_2^2 \big],
\]
with $\mathbf{x}_t = (1-t)\mathbf{x}_0 + t \mathbf{x}_1$, where $\mathbf{c}$ denotes the video, text, and reference conditions. Video features come from CLIP~\cite{clip} and Synchformer~\cite{synchformer}; audio is decoded by BigVGAN~\cite{bigvgan}.

\paragraph{Overview.}
\ModuleName{} consists of three components. First, since no existing MR-V2A training dataset provides videos paired with both ground-truth audio and corresponding reference sounds, Sec.~\ref{sec:evrefaug} mines sound events within each video and applies four variation modes to construct non-oracle reference-target pairs. Second, Sec.~\ref{sec:dualref} organizes the resulting event references into a \textbf{T}ime-\textbf{A}ligned \textbf{CO}mposite \textbf{Ref}erence (TACO-Ref), which is integrated into the DiT for reference-guided generation. Third, the injection of composite reference feature introduce cross-event acoustic-attribute leakage, Sec.~\ref{sec:refattn} employs localized reference attention together with the TACO-Ref Router (TRR) to disentangle the event-specific acoustic attributes of individual sounds.

\subsection{\AugName: Intra-Video Event Variation}
\label{sec:evrefaug}

As shown in Fig.~\ref{fig:method_overview} A, Mr-V2A requires event-level reference-target audio pairs, which are rarely available in existing datasets. Such pairs can be constructed more naturally within the same video, since multiple sound events are recorded under similar acoustic environments and recording conditions. However, directly using a target event as the reference would introduce an oracle shortcut, allowing the model to copy target-specific information rather than learn acoustic attributes. We therefore exploit intra-video event variations to construct diverse non-oracle reference-target pairs.

\paragraph{Event Discovery.}
We first identify sound events in each video. An RMS-energy detector removes long silent or low-energy regions and partitions the remaining audio into candidate event segments.  Qwen2.5-Omni~\cite{Qwen2.5-Omni} then analyzes each audio-visual segment and produces a sound category, a fine-grained description, and a confidence score. Events with low annotation confidence are discarded.

\paragraph{Timbre Variation.}
For each retained event $E_i$, we construct its reference using four variation modes:
\textbf{i) Group replacement.}
We cluster all events within the same video in the CLAP-HTSAT~\cite{clap} audio embedding space using K-Means, with the number of clusters selected automatically by the elbow criterion. Events assigned to the same cluster as $E_i$ form a candidate neighbor set $\mathcal{N}(E_i)$, from which a semantically related but non-identical event can be sampled as its timbre reference.
\textbf{ii) Speed variation.}
The selected reference event is resampled by a factor drawn from ${0.7, 0.8, 1.2, 1.5}$, introducing controlled temporal and pitch variation while retaining its underlying acoustic identity.
\textbf{iii) Mixing.}
An unrelated waveform sampled from another example in the training batch is mixed with the reference at an SNR uniformly sampled from $[-5,15]$,dB, simulating references contaminated by interfering sounds.
\textbf{iv) Acoustic degradation.}
The reference is further perturbed using additive noise, low-pass filtering, or short reverberation to emulate imperfect recording and retrieval conditions.
\textbf{v) Reference dropout.} The reference audio for each event may be replaced with silence with probability $p_{empty}$. The above augmentation modes can be combined to construct diverse reference--target pairs, with their individual contributions evaluated in Sec.~\ref{sec:ablation}.

\providecommand{\stdpm}[1]{{\scriptsize$\pm\,#1$}}

\begin{table*}[!t]
\centering
\resizebox{\textwidth}{!}{%
\begin{tabular}{l|c|ccccc|cccc}
\toprule
\multirow{2}{*}{Model} & \multirow{2}{*}{\#Params} & \multicolumn{5}{c|}{Quality / Semantic} & \multicolumn{4}{c}{Controllability (vs.\ reference)} \\
 & & FAD$_{\text{VGG}}$ $\downarrow$ & KL$_{\text{PANNs}}$ $\downarrow$ & IS$_{\text{PaSST}}$ $\uparrow$ & IB-Score $\uparrow$ & Desync $\downarrow$ & BEATs-Cos $\uparrow$ & CLAP-Cos $\uparrow$ & LSD $\downarrow$ & MCD $\downarrow$ \\
\midrule
MMAudio-L                   & 1.03B & $2.016$\stdpm{0.081} & $2.087$\stdpm{0.015} & $\underline{11.70}$\stdpm{0.09} & $\mathbf{0.3468}$\stdpm{0.0009} & $\underline{0.427}$\stdpm{0.012} & $0.6254$\stdpm{0.0017} & $0.5320$\stdpm{0.0013} & $20.06$\stdpm{0.09} & $592.8$\stdpm{2.7} \\
MMAudio-M                   & 621M & $0.911$\stdpm{0.072} & $2.082$\stdpm{0.018} & $10.03$\stdpm{0.07} & $\underline{0.3199}$\stdpm{0.0022} & $0.428$\stdpm{0.005} & $0.6207$\stdpm{0.0006} & $0.5213$\stdpm{0.0017} & $20.91$\stdpm{0.21} & $610.8$\stdpm{3.1} \\
HunyuanVideo-Foley               & 5.13B & $2.485$\stdpm{0.098} & $2.562$\stdpm{0.025} & $6.69$\stdpm{0.06} & $0.3047$\stdpm{0.0012} & $0.556$\stdpm{0.004} & $0.6192$\stdpm{0.0011} & $0.5150$\stdpm{0.0013} & $21.11$\stdpm{0.10} & $611.4$\stdpm{2.9} \\
ThinkSound & 1.28B & $2.424$\stdpm{0.087} & $2.410$\stdpm{0.018} & $8.67$\stdpm{0.06} & $0.2433$\stdpm{0.0018} & $0.670$\stdpm{0.009} & $0.6078$\stdpm{0.0007} & $0.4966$\stdpm{0.0023} & $24.15$\stdpm{0.14} & $641.3$\stdpm{5.3} \\
PrismAudio & 674M & $1.714$\stdpm{0.035} & $2.527$\stdpm{0.022} & $6.82$\stdpm{0.08} & $0.2429$\stdpm{0.0006} & $1.117$\stdpm{0.006} & $0.6055$\stdpm{0.0008} & $0.4945$\stdpm{0.0016} & $21.25$\stdpm{0.16} & $597.0$\stdpm{1.6} \\
FoleyDirector                 & 681M & $0.866$\stdpm{0.069} & $2.071$\stdpm{0.015} & $\mathbf{13.75}$\stdpm{0.28} & $0.3124$\stdpm{0.0017} & $\mathbf{0.423}$\stdpm{0.005} & $0.6221$\stdpm{0.0007} & $0.5180$\stdpm{0.0029} & $21.05$\stdpm{0.56} & $606.5$\stdpm{3.9} \\
\midrule
AC-Foley (Random)     & 1.08B & $4.657$\stdpm{1.753} & $1.772$\stdpm{0.104} & $8.75$\stdpm{0.30} & $0.2551$\stdpm{0.0165} & $0.764$\stdpm{0.047} & $0.6240$\stdpm{0.0140} & $0.4741$\stdpm{0.0323} & $26.68$\stdpm{2.43} & $519.4$\stdpm{22.0} \\
AC-Foley (Comp.)    & 1.08B & $5.746$\stdpm{1.568} & $2.273$\stdpm{0.077} & $8.55$\stdpm{0.40} & $0.2331$\stdpm{0.0096} & $0.737$\stdpm{0.016} & $0.6011$\stdpm{0.0083} & $0.4560$\stdpm{0.0225} & $27.57$\stdpm{1.72} & $597.4$\stdpm{13.9} \\
AC-Foley (Tail.)    & 1.08B & $\mathbf{0.701}$\stdpm{0.047} & $\mathbf{1.419}$\stdpm{0.019} & $7.88$\stdpm{0.15} & $0.3020$\stdpm{0.0017} & $0.545$\stdpm{0.009} & $\underline{0.6637}$\stdpm{0.0013} & $\underline{0.5767}$\stdpm{0.0023} & $\underline{17.93}$\stdpm{0.24} & $\underline{475.3}$\stdpm{2.6} \\

\midrule
\textbf{\ModuleName{} (Ours)}         & 761M & $\underline{0.839}$\stdpm{0.026} & $\underline{1.618}$\stdpm{0.018} & $10.66$\stdpm{0.19} & $0.2964$\stdpm{0.0014} & $0.476$\stdpm{0.009} & $\mathbf{0.6947}$\stdpm{0.0008} & $\mathbf{0.6286}$\stdpm{0.0013} & $\mathbf{16.92}$\stdpm{0.04} & $\mathbf{460.2}$\stdpm{1.6} \\
\bottomrule
\end{tabular}%
}
\caption{Main comparison on \EvalName. Left block: standard Foley generation quality and semantic metrics; right block: reference controllability metrics computed against the designated reference.}
\label{tab:main}
\end{table*}

\subsection{TACO-Ref Encoding and Dual-Path Injection}
\label{sec:dualref}

\paragraph{Time-Aligned Composite Reference.}
We place the perturbed reference events at their corresponding target-time positions on a shared silent waveform, forming a \textbf{T}ime-\textbf{A}ligned \textbf{CO}mposite \textbf{Ref}erence (TACO-Ref). When a reference is sampled as empty, its assigned slot remains silent and provides an empty timbre condition. A frozen BEATs audio encoder~\cite{beats} then extracts a time-indexed feature sequence $\mathbf{R}\in\mathbb{R}^{T_r\times768}$ from the complete TACO-Ref, preserving the temporal organization of the multi-event condition.

\begin{table}[!tb]
\centering
\begingroup
\setlength{\tabcolsep}{1.6pt}
\renewcommand{\arraystretch}{1.08}
\resizebox{\columnwidth}{!}{%
\begin{tabular}{@{}lcccccccc@{}}
\toprule
& \multicolumn{4}{c}{2-event} & \multicolumn{4}{c}{3-event} \\
\cmidrule(lr){2-5} \cmidrule(lr){6-9}
\multicolumn{1}{c}{Model} & Avg $\downarrow$ & Worst $\downarrow$ & Gap $\downarrow$ & DS $\downarrow$ & Avg $\downarrow$ & Worst $\downarrow$ & Gap $\downarrow$ & DS $\downarrow$ \\
\midrule
V-Foley & 672.6 & 779.8 & 214.6 & 1.320 & 718.1 & 884.7 & 336.6 & 1.252 \\
FoleyDir. & 706.1 & 799.6 & 187.0 & \textbf{0.945} & 759.2 & 916.2 & 303.9 & 0.940 \\
AC-Foley-S & 544.6 & 712.3 & 335.4 & 1.021 & 640.1 & 867.3 & 463.7 & 1.010 \\
AC-Foley-C & 608.8 & 733.2 & 248.9 & 1.008 & 654.7 & 883.2 & 478.1 & 1.038 \\
\textbf{Ours} & \textbf{512.6} & \textbf{578.0} & \textbf{130.7} & 1.006 & \textbf{504.8} & \textbf{599.0} & \textbf{185.6} & \textbf{0.932} \\
\bottomrule
\end{tabular}%
}
\endgroup
\caption{\VividBench{} multi-event reference control under two- and three-event compositions. V-Foley and FoleyDir. denote Video-Foley and FoleyDirector; AC-Foley-S and AC-Foley-C denote single-reference and composite-reference AC-Foley. Avg is the mean event-level MCD, Worst averages the worst-event MCD within each case, Gap denotes the within-case MCD range, and DS denotes DeSync. }
\label{tab:main_split}
\end{table}

\paragraph{Dual-path Injection.}
In Fig.~\ref{fig:method_overview} B, We inject $\mathbf{R}$ through Dual-pathways. The local pathway retains its temporal resolution and provides event-specific reference cues to the audio latent, whereas the global pathway aggregates the complete TACO-Ref into a shared acoustic-context condition.

\paragraph{Local: time-aligned reference injection.}
We project the TACO-Ref features into the DiT hidden space and introduce an additional Reference Attention module after the original latent-video-text joint attention. Given the audio latent $\mathbf{h}_a$, we obtain the query $\mathbf{Q}_a=\operatorname{Proj}_{q}(\mathbf{h}_a)$ and compute the reference key and value from $\mathbf{R}$:
\begin{align*}
\mathbf{A}_{\mathrm{ref}} &= \operatorname{softmax}\!\left(\frac{\mathbf{Q}_a\mathbf{K}_r^{\mathsf T}}{\sqrt{d}}\right)\mathbf{V}_r,\\
(\mathbf{K}_r,\mathbf{V}_r) &= \operatorname{Proj}_{kv}(\mathbf{R}).
\end{align*}
The resulting update is added to the audio latent:
$$
\mathbf{h}_a^{\prime}=\mathbf{h}_a+\gamma_{\mathrm{ref}}\mathbf{A}_{\mathrm{ref}}\mathbf{W}_o,
$$
where $\gamma_{\mathrm{ref}}$ is initialized to zero. Retaining the temporal structure of $\mathbf{R}$ enables the subsequent localized attention and TRR modules to associate each audio-latent region with its designated reference event.

\paragraph{Global: acoustic-context injection.}
In parallel, an Audio Q-Former aggregates the complete TACO-Ref using $N_q=32$ learnable queries and $L=8$ layers:
$$
\mathbf{Q}=\mathrm{QFormer}(\mathbf{R})\in\mathbb{R}^{N_q\times d_h},\qquad \bar{\mathbf{q}}=\frac{1}{N_q}\sum_{i=1}^{N_q}\mathbf{Q}_i.
$$
We project $\bar{\mathbf{q}}$ into the pooled video--text conditioning space:
$$
\tilde{\mathbf{c}}_{v,t}^{\mathrm{pool}}=\mathbf{c}_{v,t}^{\mathrm{pool}}+\mathbf{W}_q\bar{\mathbf{q}}.
$$
The updated global condition is injected through the DiT's adaptive normalization layers, allowing the model to modulate acoustic properties shared across event slots, such as background texture and overall spectral characteristics.

\begin{table}[!tb]
\centering
\resizebox{\columnwidth}{!}{%
\begin{tabular}{cccccc}
\toprule
Exp & Local & Global & FAD$_{\text{VGG}}$ $\downarrow$ & KL $\downarrow$ & BEATs-Cos $\uparrow$ \\
\midrule
\textcircled{1}       & \checkmark & --             & 0.977 & 1.98 & 0.652 \\
\textcircled{2}    & --         & q-former         & 1.614 & 2.14 & 0.669 \\
\textcircled{3}    & \checkmark & mean & 1.009 & 1.91 & 0.652 \\
\textcircled{4}  & \checkmark & q-former & \textbf{0.839} & \textbf{1.62} & \textbf{0.695} \\
\bottomrule
\end{tabular}%
}
\caption{Ablation on Dual-path injection on VGGSound-Ref.}
\label{tab:abl_arch}
\end{table}

\subsection{Multi-Timbre Disentanglement}
\label{sec:refattn}

When TACO-Ref contains references to multiple sound events, unrestricted reference attention may mix features across event boundaries, leading to unintended leakage of acoustic attributes between events. We therefore disentangle event-specific acoustic attributes: localized reference attention restricts each audio query to a nearby temporal neighborhood, while the TACO-Ref Router establishes explicit event-to-slot correspondence by routing each event-specific reference response to its assigned audio region according to soft temporal slot masks.

\paragraph{Localized Reference Attention.}
As shown in Fig.~\ref{fig:method_overview} C.1, allowing each audio latent to attend globally over the entire TACO-Ref may expose it to reference features from temporally unrelated events. We therefore impose a local attention window that aligns reference conditioning with the corresponding audio region. Specifically, for an audio latent query at position $i$, only reference keys and values within $\pm w$ frames are accessible:

\[
M_{ij}=
\begin{cases}
0, & |i-j|\leq w,\\
-\infty, & |i-j|>w.
\end{cases}
\]

The localized reference attention is then computed as
\[
\mathbf{A}_{\mathrm{loc}}
=
\operatorname{softmax}\!\left(
\frac{\mathbf{Q}_a\mathbf{K}_r^{\mathsf T}}{\sqrt{d}}+\mathbf{M}
\right)\mathbf{V}_r.
\]

During training, we use a fixed window size of $w{=}50$, corresponding to approximately $\pm0.5$s. At inference time, $w$ can be adjusted without retraining to control the granularity of attribute disentanglement: a smaller window enforces stronger temporal isolation, whereas a larger window allows broader reference context.

\paragraph{TACO-Ref Router (TRR).}
Localized reference attention suppresses interference from temporally distant reference frames, but a fixed local window can still include concurrent or closely spaced events. We therefore introduce the TACO-Ref Router (TRR) to impose an explicit event-to-slot correspondence between reference features and audio latent regions in Fig.~\ref{fig:method_overview} C.2. For each event $k$, TRR constructs an event-specific TACO-Ref bank by placing only its reference clip at the assigned target span and encoding the resulting per-event canvas, yielding $\mathbf{R}_k$. Given the audio latent query $\mathbf{Q}_a$, TRR computes an event-specific reference attention:
$$
(\mathbf{K}_k,\mathbf{V}_k)=\operatorname{Proj}_{kv}(\mathbf{R}_k),
\qquad
\mathbf{A}_k=\operatorname{Attn}(\mathbf{Q}_a,\mathbf{K}_k,\mathbf{V}_k).
$$
Let $\mathbf{m}^{a}_k(t)$ denote the audio-latent slot mask for event $k$, obtained by projecting its target time span onto the latent time grid. The event-specific responses are then fused with the slot masks:
$$
\mathbf{A}_{\mathrm{TRR}}(t)
=
\sum_{k=1}^{K}
\mathbf{m}^{a}_k(t)\mathbf{A}_k(t).
$$
The masks are soft around event boundaries, allowing smooth transitions between adjacent slots. In this way, localized attention provides fine-grained temporal locality within each event, while TRR biases each audio region toward its assigned reference event and reduces leakage of acoustic attributes to unrelated regions. \textbf{More Details in appendix}.

\begin{figure}[!tb]
\centering
\includegraphics[width=\columnwidth]{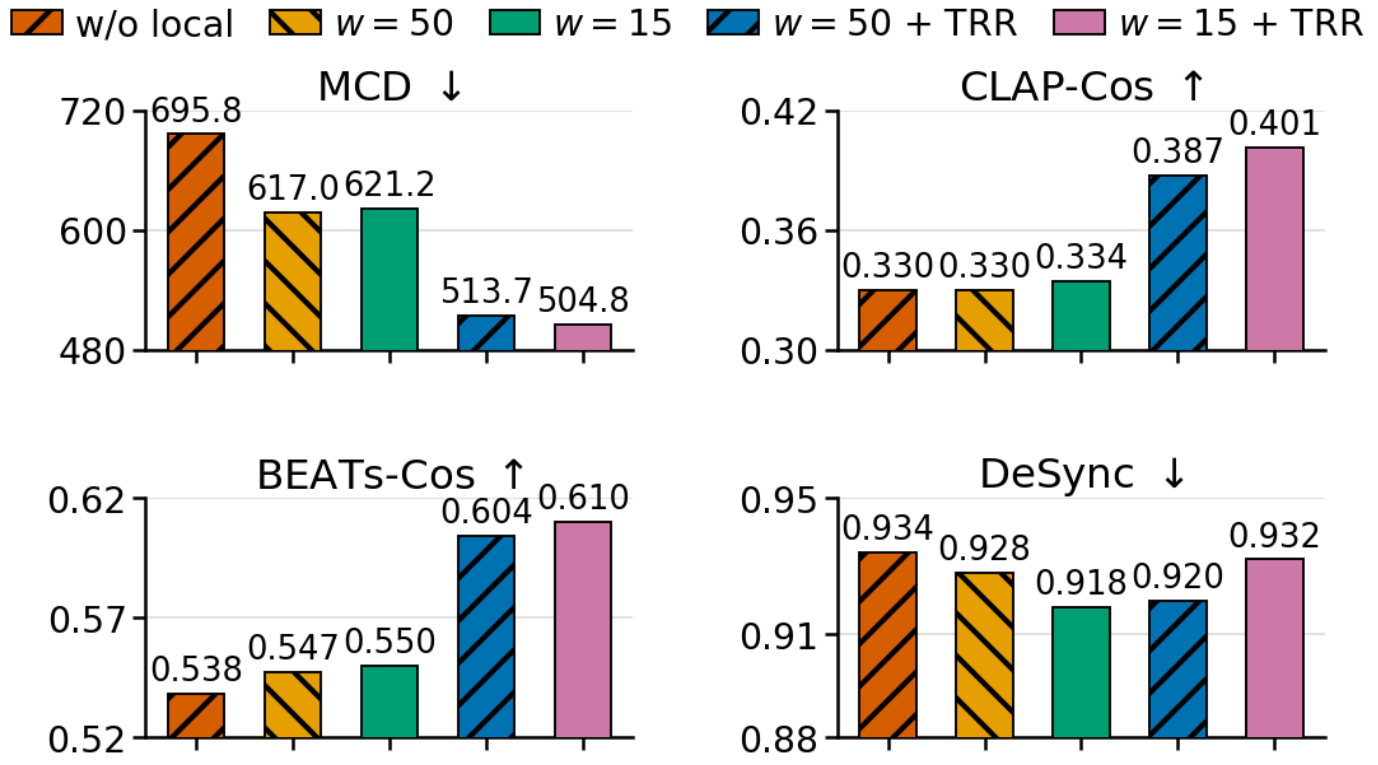}
\caption{VividBench three-event ablation of localized reference attention and the TACO-Ref Router (TRR).}
\label{fig:abl_attn}
\end{figure}

\section{Experiments}
\label{sec:experiments}

\subsection{Datasets, Metrics, and Baselines}
\label{sec:exp_setup}

\paragraph{Training data.}
We train on VGGSound~\cite{vggsound}, AudioCaps~\cite{audiocaps}, and an internal long-video corpus. Each corpus is converted into event-level reference-target examples for \AugName{}, and the implementation details are provided in the supplementary material.

\paragraph{Benchmarks.}
\textbf{1) \EvalName{}.} It is built from the held-out VGGSound test split with methods described in Sec.~\ref{sec:evrefaug}, using \textit{non-overlapping intra-video event segments as references}, avoiding the oracle shortcut of using the target event itself as the reference. It evaluates reference-conditioned Foley generation under the VGGSound distribution and supports both standard generation metrics and event-level acoustic identity metrics. 
\textbf{2) \VividBench{}.}
To evaluate controllability under more realistic reference conditions, we manually select 50 videos and annotate their sound-event categories (with three timbres per video) and temporal spans. For each event, we retrieve an external audio reference from Freesound~\cite{freesound}. See supplementary material for details.

\paragraph{Metrics.}
We report FAD$_{\text{VGG}}$~\cite{FAD,vggish}, KL$_{\text{PANNs}}$~\cite{audioldm, panns}, IS$_{\text{PaSST}}$~\cite{isscore,passt}, IB-Score~\cite{imagebind}, and Desync~\cite{synchformer} for generation quality and video-event alignment on \EvalName{}. Reference controllability is measured with BEATs-Cos, CLAP-Cos, LSD (\emph{Log-Spectral Distance}, a low-level spectral deviation between the generated audio and the reference audio), and MCD used in AC-Foley; on \VividBench{}, the acoustic metrics are computed at the event level using the annotated target spans.

\begin{figure*}[t]
    \centering
    \includegraphics[width=0.82\textwidth]{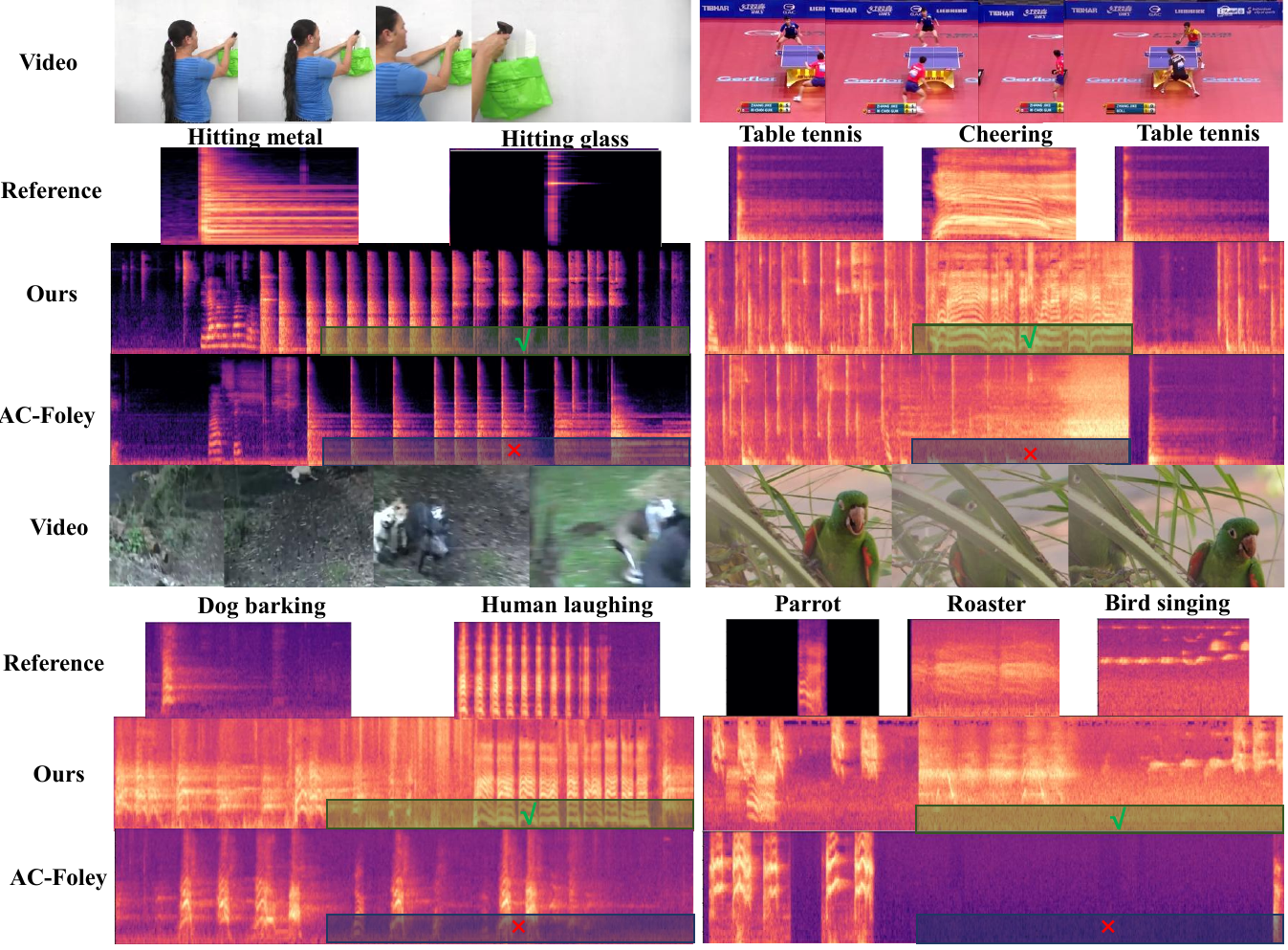}
    \caption{\textbf{Qualitative comparison on \VividBench{}.} From top to bottom, the rows show video frames, event-level reference spectrograms, outputs from VividAudio, and outputs from AC-Foley.}
    \label{fig:vividbench_visualization}
\end{figure*}

\paragraph{Baselines.}
We compare the following methods: MMAudio-L/M, HunyuanVideo-Foley, ThinkSound, PrismAudio~\cite{prismaudio}, FoleyDirector, and AC-Foley. We additionally compare with Video-Foley on \VividBench{}. \textit{Tri-Ergon and FoleyGenEx are not included because their official implementations are not publicly available.}
For a \textbf{fair comparison}, we evaluate AC-Foley under multiple reference-construction settings. On \EvalName{}, we report 1) Default setting, denoted as \textit{Tail}, which uses the final 2s of each 10s ground-truth audio clip as the reference condition. 2) \textit{random.}, which randomly selects one event reference, 3) \textit{Comp.}, which concatenates all references in temporal order into a composite reference. On \VividBench{}, ground-truth audio is unavailable, making the \textit{tail} setting inapplicable. We evaluate two settings: \textit{Sing.}, which uses only the reference associated with the first event, and \textit{Comp.}, which concatenates all event references in temporal order.

\paragraph{Implementation.}
The backbone is a medium MMAudio DiT in the latent space of a 44kHz audio autoencoder. Training is conducted on eight 40GB GPUs, using a learning rate of $1\times10^{-5}$, a cosine learning-rate schedule, an effective batch size of $48$, and $4\times10^{5}$ SFT steps. Inference uses a guidance scale of $s=4.5$. All experiments are conducted with five random seeds: 42, 43, 44, 666, and 888. We set the Q-former with $N_q=32$, $L=8$.

\subsection{Main Comparison}
\label{sec:exp_main}

\paragraph{\EvalName{}.}
In Table~\ref{tab:main}, \ModuleName{} achieves the strongest reference controllability, with a BEATs-Cos of $0.695$, a CLAP-Cos of $0.629$, an LSD of $16.92$, and an MCD of $460.2$. Meanwhile, it remains competitive in generation quality, with an $\text{FAD}_{\text{VGG}}$ of $0.839$ and a $\text{KL}_{\text{PANNs}}$ of $1.618$, while retaining temporal alignment with Desync of $0.476$. AC-Foley achieves competitive performance only under its default reference setting, indicating limited generalization to event-level reference control and highlighting the advantage of our method in handling multiple distinct timbres.

\paragraph{VividBench.}
Table~\ref{tab:main_split} reports results for the two- and three-event settings using event-level MCD. \textbf{1) Strongest controllability.} \ModuleName{} achieves the lowest average MCD in both settings ($512.6/504.8$). \textbf{2) Robust multi-event timbre control.} It also reduces the worst-event MCD to $578.0/599.0$ and the lowest within-case MCD gap ($130.7/185.6$), indicating consistent control across events rather than gains concentrated on a single easy event. In contrast, AC-Foley exhibits substantially larger MCD gaps, reflecting its limited ability to control multiple event-specific timbres. \textbf{3) Effective temporal alignment.} \ModuleName{} achieves the best three-event Desync score of $0.932$, while FoleyDirector performs best in the two-event setting with $0.945$.

\paragraph{Qualitative analysis.}
Figure~\ref{fig:vividbench_visualization} shows two representative \VividBench{} cases. \textbf{1) Multi-event controllability}  Across both the animal-event and heterogeneous outdoor-scene cases, \ModuleName{} more consistently preserves the acoustic characteristics of the designated references for successive, repeated, and heterogeneous events. In contrast, AC-Foley exhibits weaker event-wise correspondence, particularly within the highlighted intervals. \textbf{2) Not a simply copy of reference.} The generated spectrograms is not simply reproducing or concatenating the reference audios. This observation also supports the effectiveness of our non-oracle reference construction.

\begin{table}[!tb]
\centering
\resizebox{\columnwidth}{!}{%
\begin{tabular}{lcccc}
\toprule
Sampling Prob. & KL $\downarrow$ & IB-Score $\uparrow$ & BEATs-Cos $\uparrow$ & LSD $\downarrow$ \\
\midrule
$(1,0,0,0,0)$  (group)               & 1.77 & 0.2740 & 0.682 & 17.43 \\
$(0,1,0,0,0)$  (speed)               & 1.65 & 0.2775 & 0.697 & 17.06 \\
$(0,0,1,0,0)$  (mix)                 & 1.96 & 0.2347 & 0.729 & 16.51 \\
$(0,0,0,1,0)$  (deg)                 & 1.97 & 0.1864 & \textbf{0.756} & \textbf{15.87} \\
$(0,0.8,0,0,0.2)$                    & 1.64 & 0.2842 & 0.693 & 17.27 \\
$(0.2,0.2,0.2,0.2,0.2)$              & 1.80 & 0.2838 & 0.697 & 17.36 \\
$(0.4,0.4,0,0,0.2)$                  & \textbf{1.62} & \textbf{0.2964} & 0.695 & 16.92 \\
$(0.4,0.4,0,0.1,0.1)$                & 1.73 & 0.2852 & 0.704 & 16.79 \\
$(0.4,0.4,0.1,0,0.1)$                & 1.77 & 0.2828 & 0.699 & 17.08 \\
\bottomrule
\end{tabular}%
}
\caption{Ablation on \AugName. The Sampling Prob is $(p_{\text{group}}, p_{\text{speed}}, p_{\text{mix}}, p_{\text{deg}}, p_{\text{empty}})$, corresponding to the four variation modes and empty dropout.}
\label{tab:abl_aug}
\end{table}

\subsection{Ablation Studies}
\label{sec:ablation}

\paragraph{TACO-Ref feature injection.}
Table~\ref{tab:abl_arch} evaluates the local and global TACO-Ref injection paths. Adding the pooled mean condition improves KL from $1.98$ to $1.91$, with comparable FAD$_{\text{VGG}}$ ($0.977$ vs.\ $1.009$). Replacing mean pooling with the Audio Q-Former further improves FAD$_{\text{VGG}}$ to $0.839$ and BEATs-Cos from $0.652$ to $0.695$. More ablation on Q-Former are provided in the supplementary material.

\paragraph{Multi-timbre disentanglement.}
Fig.~\ref{fig:abl_attn} evaluates localized reference attention and TRR on the three-event split of \VividBench{}. First, removing the local attention window results in the highest MCD and the lowest CLAP-Cos and BEATs-Cos scores. Introducing localized reference attention substantially improves timbre similarity, while TRR provides further gains without noticeably degrading audio--video alignment. Second, we study the effect of the inference window size $w$ while fixing the training window size to 50. The results show that localized attention is the primary source of improvement, whereas adjusting $w$ at inference yields only modest additional gains.

\begin{figure}[!tb]
\centering
\includegraphics[width=\columnwidth]{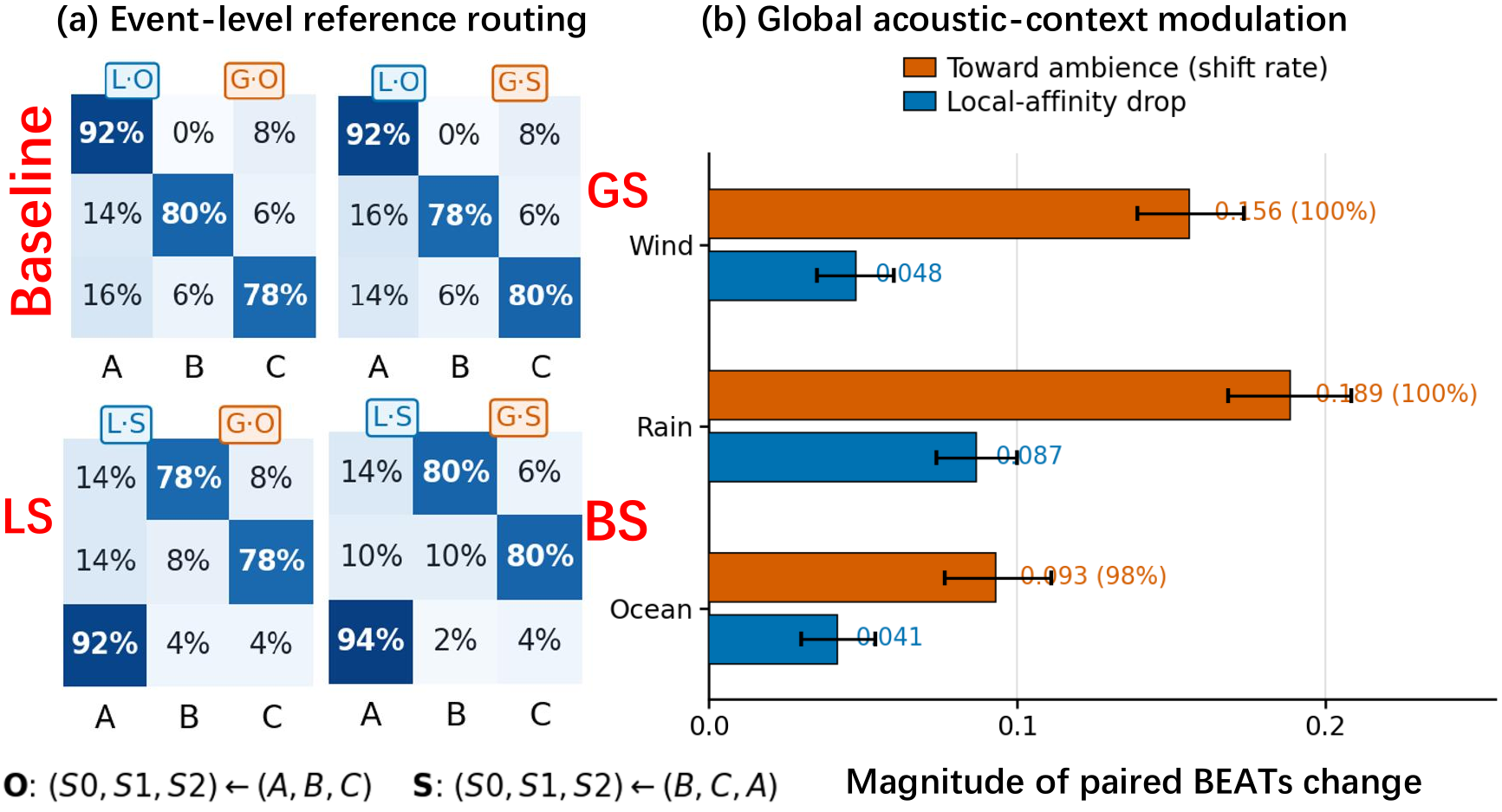}
\caption{Causal pathway analysis on the VividBench three-event split. \textbf{(a)} Under cyclic reference reassignment, $\mathrm{O}/\mathrm{S}$ denote original/swapped slot assignments, and $\mathrm{L}/\mathrm{G}$ denote local/global pathways. Heatmaps report the nearest reference event by MCD for each generated slot. \textbf{(b)} With local references fixed, global ambience injection is measured by BEATs shift toward the ambience and local-reference affinity loss. Error bars show bootstrap $95\%$ confidence intervals; parentheses indicate ambience-following rates.}
\label{fig:path_roles}
\end{figure}

\paragraph{Pathway analysis.}
We analyze the roles of the local and global reference conditions through paired counterfactual interventions on the three-event split of \VividBench{}. For each case, we cyclically reassign the same three reference events from the original slot order $(A,B,C)$ to $(B,C,A)$ in the global condition only (GS), the local condition only (LS), or both conditions (BS), while fixing the video, text prompt, random seed, reference set, and target event intervals. In Figure~\ref{fig:path_roles}(a), the baseline exhibits the expected diagonal pattern, meaning each generated slot is closest to its originally assigned reference. \textit{Swapping only the global condition} leaves this diagonal structure \textit{largely unchanged}, indicating that the global pathway alone \textit{\textbf{does not determine}} which event-level timbre is selected for each slot. In contrast, swapping the local condition or swapping both pathways \textit{redirects the nearest-reference pattern} to the cyclic off-diagonal structure. 

We then fix all local event references and replace only the global condition with an $8$\,s recording of wind, rain, or ocean ambience. Figure~\ref{fig:path_roles}(b) shows that every ambience produces a substantially larger shift toward the injected global context than the accompanying loss of local-reference affinity. The generated soundtrack moves toward the injected ambience in $98\%$--$100\%$ of cases. These results identify the global condition as a clip-level acoustic-context pathway that controls shared background texture and spectral characteristics without overriding event-level assignment. 

\paragraph{Intra-Video Event Variation.}
In Table~\ref{tab:abl_aug}, Group replacement and speed variation improve timbre consistency while largely preserving generation quality. In contrast, mixing and acoustic degradation provide larger consistency gains but introduce noticeable quality degradation. Randomly dropping a subset of event references helps preserve the model's original quality and semantic distributions by preventing excessive reliance on reference conditions. Combining complementary modes increases reference diversity and achieves a better balance between audio quality and timbre consistency.

\subsection{MLLM Preference Evaluation}
\begin{figure}[!t]
\centering
\includegraphics[width=0.92\columnwidth]{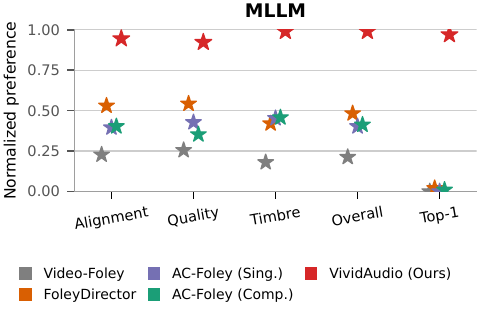}
\caption{MLLM preference comparison on \VividBench{}.  Mean ranks are converted to a higher-is-better preference score as $(5-\text{rank})/4$, while top-1 rates remain in $[0,1]$.}
\label{fig:human_omni}
\end{figure}

We conduct an MLLM-based preference evaluation on \VividBench{} with Gemini2.5-pro~\cite{gemini25} in Figure~\ref{fig:human_omni}. Specifically, we evaluate 100 cases from 50 videos, with 50 cases each under the two-event and three-event settings generated by the five methods reported in Tab.~\ref{tab:main_split}. The method identities are anonymized and the candidates are presented in randomized order, after which Gemini independently ranks them in terms of video--event alignment, audio quality, reference-timbre controllability, and overall preference. \ModuleName{} achieves the best mean rank across all four criteria. \textbf{Details are provided in the appendix.}

\section{Conclusion}
\label{sec:conclusion}

We formulated multi-event reference-conditioned V2A (Mr-V2A) and presented \ModuleName{} for controlling the acoustic identities of multiple events within a synchronized soundtrack. To address scarce event-level supervision, \AugName{} constructs non-oracle reference-target pairs from events mined within the same video. TACO-Ref then organizes the references on a shared temporal canvas and injects frozen BEATs features through local and global pathways, while localized reference attention and TRR reduce cross-event attribute leakage through explicit event-to-slot routing.
Experiments on \EvalName{} and \VividBench{} demonstrate stronger and more balanced reference controllability than conventional V2A baselines while maintaining competitive Foley generation quality and temporal alignment. Ablations verify the contributions of dual-path conditioning, localized attention, TRR, and intra-video event variation.
These results establish Mr-V2A as a step toward more flexible user-directed audio creation. Future work includes extending control beyond audio exemplars, and scaling event-specific reference conditioning to longer and more complex videos.

\section*{Acknowledgements}
This work was supported by the Fundamental Research Funds for the Central Universities (226-2025-00080), the National Natural Science Foundation of China (62293554) and in part by the Natural Science Foundation of Zhejiang Province (LDT23F02023F02).
\bigskip
\bibliography{aaai2027}

\clearpage
\appendix
\section{Supplementary Data Construction Details}
\label{sec:supp_data_pipeline}

This section describes the implementation details of the data pipeline introduced in Sec.~3.2 of the main paper. We focus on corpus processing, event extraction, graph serialization, and training triplet export; the motivation for intra-video event variation augmentation and the reference perturbation strategy are described in the method section of the main paper.

\subsection{Corpora and Event Extraction}
\label{subsec:segmentation}
We process VGGSound~\cite{vggsound}, AudioCaps~\cite{audiocaps}, and an internal corpus. Source clips are first segmented with an RMS-energy event detector. An adaptive energy threshold removes long silent or low-energy regions, and each retained event is cropped to a duration between $0.2$\,s and $8.0$\,s.

Each audio-visual crop is then passed to Qwen2.5-Omni~\cite{Qwen2.5-Omni}, which outputs structured metadata containing a sound category, a fine-grained description, and a confidence score. Events with confidence below $0.6$ are discarded. The remaining records retain the source-video identifier, temporal boundaries, audio path, visual crop, and annotation fields required by the graph construction and triplet generation stages.

\subsection{Feature Extraction and Graph Serialization}
\label{subsec:clustering}
For every retained event, we resample the audio to $48$\,kHz and extract the $512$-dimensional global pooled representation from CLAP-HTSAT~\cite{clap}. K-Means is applied independently within each source video. We evaluate $K\in\{2,\ldots,8\}$ and select the elbow point using the local change in clustering inertia. This produces a group identifier for every event without requiring a global category-specific clustering model.

After clustering, each event receives two graph fields. The first is a group identifier, which marks the intra-video acoustic cluster to which the event belongs. The second is a neighbor list containing the indices of other events assigned to the same cluster. These neighbors define the candidate pool for non-oracle reference construction. If clustering fails, or if an event forms a singleton cluster, we store \texttt{group=-1} or an empty neighbor list. This makes the annotation format robust to sparse or ambiguous videos and avoids forcing every event to have a same-group partner.

\subsection{Neighbor Records and Triplet Export}
\label{subsec:retrieval_graph}
For each event $E_i$, the pipeline serializes its group identifier and the indices of all group-associated events as a neighbor list:

\begin{verbatim}
{
  "label":     "foley",
  "desc":      "wooden-floor steps",
  "group":     2,
  "neighbors": [4, 7, 12]
}
\end{verbatim}

The neighbor lists are consumed by the perturbation process described in Sec.~3.2 of the main paper. Concretely, the exporter constructs event references from a perturbation family including \emph{same-group replacement}, \emph{speed variation}, \emph{mixing}, \emph{acoustic degradation}, and an optional \emph{empty reference} condition. Same-group replacement samples a non-identical event from the neighbor list when a valid same-group partner exists. In the empty-reference case, no waveform is pasted into the corresponding slot, so the slot remains silent in TACO-Ref and acts as an explicit zero-reference condition rather than a missing-data error.

For bookkeeping, the exported JSONL keeps both the audio paths and the temporal spans used on the reference side and the generation side. A typical record contains \texttt{rag\_audio\_list}, \texttt{rag\_seg\_in\_ref}, and \texttt{rag\_seg\_in\_gen} for the event-level references, together with \texttt{inp\_audio\_list}, \texttt{inp\_seg\_in\_ref}, and \texttt{inp\_seg\_in\_gen} for the optional inpainting/background regions. In the simplest exporter, if no valid same-group neighbor exists, the system falls back to self-reconstruction for that event so that the sample remains usable; the subsequent perturbation stage can still convert that reference into a non-oracle or empty condition. The exporter then writes the composite reference waveform and the resulting $(V,a_r,a_1)$ paths into the training JSONL files. The same metadata format supports reference feature extraction, TACO-Ref construction, and reproducible sample-level bookkeeping.

\section{Background on Diffusion-Based Audio Generation}
\label{sec:supp_diffusion_audio_background}

Modern audio generation systems often follow the same high-level design as image latent diffusion: they learn a generative process in a compact latent space and decode the generated latent back to waveform audio. Early text-to-audio systems apply diffusion either on spectrogram-like representations or learned audio latents, with AudioLDM showing that a latent diffusion model can synthesize realistic audio from text conditions~\cite{audioldm,audioldm2}. Video-to-audio models extend this formulation by adding visual and temporal conditions, so the generated soundtrack must be both acoustically plausible and synchronized with the input video~\cite{difffoley,frieren,mmaudio}.

\subsection{Latent Audio Representation}

Direct waveform generation is expensive because audio has a high sampling rate and fine temporal structure. A common solution is to train an audio autoencoder that maps waveform audio $a$ to a lower-rate latent sequence $\mathbf{x}$ and then reconstructs the waveform from the generated latent. In this setting, the diffusion or flow model operates on $\mathbf{x}$ rather than raw samples. This reduces sequence length while retaining perceptually important spectral and temporal information. In \ModuleName{}, the DiT backbone operates in the latent space of a $44$\,kHz audio autoencoder, and the final waveform is recovered with a neural vocoder following the common audio-decoding practice used in neural vocoders such as BigVGAN~\cite{bigvgan}.

\subsection{Diffusion and Rectified Flow}

Diffusion models learn to reverse a corruption process that gradually maps clean data to noise. Rectified flow uses a closely related continuous-time view but trains the network to predict a velocity field along a simple path between a noise sample $\mathbf{x}_0$ and a data latent $\mathbf{x}_1$. With a linear interpolation
\[
\mathbf{x}_t=(1-t)\mathbf{x}_0+t\mathbf{x}_1,\qquad t\in[0,1],
\]
the training target is the displacement $\mathbf{x}_1-\mathbf{x}_0$. A conditional model $v_\theta(\mathbf{x}_t,t,\mathbf{c})$ is optimized to predict this velocity given the current latent, time, and conditioning signal:
\[
\mathcal{L}_{\mathrm{FM}}=\mathbb{E}_{\mathbf{x}_0,\mathbf{x}_1,t,\mathbf{c}}
\left\|v_\theta(\mathbf{x}_t,t,\mathbf{c})-(\mathbf{x}_1-\mathbf{x}_0)\right\|_2^2.
\]
At inference time, generation starts from noise and integrates the learned velocity field toward the data distribution. Compared with the classical denoising interpretation, this formulation makes it convenient to express the model as a conditional trajectory from noise to audio latent, which is the formulation used by recent efficient V2A systems~\cite{frieren,mmaudio}.

\subsection{Conditioning for Audio and V2A}

For text-to-audio generation, the condition $\mathbf{c}$ is often a text embedding. For V2A, $\mathbf{c}$ must also include visual semantics and temporal alignment cues. A video-conditioned audio model therefore needs to decide what sound should occur, when it should occur, and what acoustic texture it should have. Existing V2A models mainly use global video/text conditions or synchronized visual features to guide the latent generator~\cite{difffoley,mmaudio}. \ModuleName{} adds event-level reference audio to this conditioning interface. The video and text conditions provide semantic and temporal grounding, while TACO-Ref provides reference acoustic identity. The local and global reference pathways described in the main paper are therefore not separate generators; they are additional conditioning routes that bias the same latent flow model toward the desired event-level and clip-level acoustic attributes.

\section{Supplementary Method Details}
\label{sec:supp_method_details}

This section expands the implementation details of \ModuleName{} that are abbreviated in the main paper. We focus on how event-level references are converted into TACO-Ref, how the local and global pathways consume the reference condition, and how event-level acoustic identities are separated when multiple references are present.

\subsection{TACO-Ref Construction}
\label{sec:supp_tacoref}

For a training or evaluation sample with $K$ annotated target events, we represent the reference set as $\mathcal{R}=\{(a^r_k,\tau_k)\}_{k=1}^{K}$, where $a^r_k$ is the reference waveform assigned to event $k$ and $\tau_k=[s_k,e_k]$ is the target span in the generated soundtrack. TACO-Ref is initialized as a silent waveform with the same duration as the target audio. Each reference waveform is cropped or stretched according to the event construction rule and then placed at the temporal location $\tau_k$. If multiple events are present, all event references are placed into the same canvas, preserving their intended target-time correspondence.

When an event is sampled with an empty reference condition, its assigned slot remains silent rather than being filled with a special token or a learned embedding. This gives the model an explicit zero-timbre condition for that event span and prevents it from assuming that every event is always supported by a valid acoustic exemplar. Empty reference conditions are sampled at the event level, so one event in a multi-event example can be unconditioned while the remaining events still receive valid references.

\subsection{Reference Perturbation Details}
\label{sec:supp_ref_perturb}

\AugName{} uses four reference-side acoustic variation modes to avoid target-copying shortcuts and to increase the diversity of acoustic conditions. Group replacement samples a non-identical neighbor from the same intra-video acoustic group. Speed variation modifies the selected reference with factors from $\{0.7,0.8,1.2,1.5\}$. Mixing adds an unrelated waveform at a randomly sampled SNR to simulate contaminated retrieved references. Acoustic degradation applies simple signal perturbations such as additive noise, low-pass filtering, and short reverberation. In addition, a small portion of event references are left empty as described above. Together, these operations define the reference perturbation family used to diversify event-level reference conditions during training, and the mixture probabilities in the ablation table summarize their overall sampling recipe.

\subsection{Dual-Path Reference Injection}
\label{sec:supp_dual_path}

The local pathway uses the time-indexed TACO-Ref feature sequence and injects it through reference attention. Since the reference canvas is aligned with the target audio timeline, a query from an audio latent region can attend to reference frames near the same target time. This pathway provides event-specific acoustic identity cues and is responsible for fine-grained timbre control within each event span.

The global pathway summarizes the complete TACO-Ref with an Audio Q-Former and injects the resulting pooled condition through adaptive normalization. This pathway does not preserve event-level slot identity. Instead, it provides clip-level acoustic context, such as recording texture, background ambience, and broad spectral characteristics shared across the soundtrack. The pathway-role analysis in the main paper empirically supports this separation: local reference swapping changes event-level identity, whereas global ambience replacement shifts the clip-level acoustic context while largely preserving local event assignment.

\subsection{Localized Reference Attention and TRR}
\label{sec:supp_trr}

Global reference attention over the entire TACO-Ref can expose an audio query to unrelated reference events, especially in multi-event cases. Localized reference attention therefore masks reference frames outside a temporal window around each audio-latent query. The mask is applied before the attention softmax, so inaccessible reference positions contribute zero probability after normalization. This constraint reduces leakage from temporally distant references while preserving differentiable attention within the allowed window.

The TACO-Ref Router (TRR) further handles cases where nearby or adjacent events fall within the same local attention window. For each event $k$, TRR builds an event-specific reference canvas that contains only the reference of that event at its assigned target span. The model computes an attention response $\mathbf{A}_k$ from each event-specific canvas and then fuses these responses using soft slot masks $\mathbf{m}^{a}_k(t)$ projected onto the audio latent time grid:
\[
\mathbf{A}_{\mathrm{TRR}}(t)=\sum_{k=1}^{K}\mathbf{m}^{a}_k(t)\mathbf{A}_k(t).
\]
The masks are softened around event boundaries to avoid abrupt transitions. In practice, TRR acts as a soft event-to-slot bias rather than a hard routing rule: each target region is encouraged to use the reference response assigned to the corresponding event, while boundary frames can smoothly interpolate between adjacent slots.

\section{Supplementary Experimental Details}
\label{sec:supp_experimental_details}

This section provides implementation and evaluation details that are abbreviated in the main paper.

\subsection{Training and Inference Configuration}
\label{sec:supp_config}

We train the medium-scale DiT backbone on VGGSound, AudioCaps, and an internal corpus after converting the clips into event-level reference-target examples. The model is optimized with SFT for $4\times 10^{5}$ steps using a learning rate of $1\times10^{-5}$, cosine scheduling, and an effective batch size of $48$. Audio is represented at $44$\,kHz. Unless otherwise specified, inference uses guidance scale $s=4.5$.

For multi-event reference conditioning, the event references are placed into their designated temporal slots to form TACO-Ref. We resample TACO-Ref to $16$\,kHz and extract $768$-dimensional frame-level features using the frozen BEATs iter3+ model pretrained on AS2M~\cite{beats}; the encoder remains fixed during both training and inference. Reference Attention operates on this time-aligned representation, while the Audio-Q-Former-aggregated condition is injected through the adaptive normalization pathway. The localized attention window is trained with $w=50$ frames and can be adjusted at inference; unless otherwise specified, the final \VividBench{} configuration uses $w=15$ for stronger event-level isolation. The TACO-Ref Router uses the event slot masks to combine event-specific reference responses.

\begin{table}[!tb]
\centering
\resizebox{\columnwidth}{!}{%
\begin{tabular}{lccc}
\toprule
$(N_q, L)$ & FAD$_{\text{VGG}}$ $\downarrow$ & KL$_{\text{PANNs}}$ $\downarrow$ & BEATs-Cos $\uparrow$ \\
\midrule
$(16, 4)$ & 1.017 & 1.773 & 0.676 \\
$(32, 4)$ & 0.933 & 1.727 & 0.681 \\
$(32, 8)$ (selected) & \textbf{0.839} & \textbf{1.618} & \textbf{0.695} \\
$(64, 8)$ & 0.912 & 1.749 & 0.705 \\
\bottomrule
\end{tabular}%
}
\caption{Audio Q-Former size comparison for pooled TACO-Ref conditioning. $(32,8)$ gives the best overall trade-off among the tested configurations, so we use it in the final model.}
\label{tab:qformer_sweep}
\end{table}

Table~\ref{tab:qformer_sweep} shows that increasing the Audio Q-Former capacity from $(16,4)$ to $(32,4)$ consistently improves FAD$_{\text{VGG}}$, KL$_{\text{PANNs}}$, and BEATs-Cos, indicating that a modest increase in query count is beneficial for summarizing the composite reference. Further deepening the module to $(32,8)$ yields the best overall trade-off and is therefore adopted in the final model. Increasing the query count again to $(64,8)$ does not bring additional gains and slightly weakens all three metrics, suggesting that simply enlarging the pooled global branch is less effective than keeping a moderate query budget with sufficient depth.

\subsection{\EvalName{} and Baseline Protocols}
\label{sec:supp_evalname_protocol}

\EvalName{} is built from the held-out VGGSound test split using the same event-discovery and intra-video reference-construction pipeline described in the main paper. For each target event, we select a \emph{non-overlapping} event segment from the same source video as the reference condition, rather than reusing the target event itself. This preserves acoustic relatedness within the same recording while avoiding the oracle shortcut of conditioning on the exact target segment. The resulting benchmark remains in-domain with respect to VGGSound, but explicitly evaluates whether a model can follow an event-level acoustic reference rather than simply generate plausible generic Foley.

On \EvalName{}, we report both full-clip generation metrics and reference-following metrics. FAD$_{\text{VGG}}$, KL$_{\text{PANNs}}$, IS$_{\text{PaSST}}$, IB-Score, and Desync are computed on the generated clips following the standard protocol of each metric. Reference controllability is evaluated with BEATs-Cos, CLAP-Cos, LSD, and MCD between the generated audio and the assigned reference condition. On \VividBench{}, these controllability metrics are evaluated at the event level using the manually annotated target spans.

For metric computation, \EvalName{} uses the standard av-benchmark pipeline for full-clip metrics and a separate reference-following evaluation for event-conditioned metrics. On \VividBench{}, for each event we crop the generated waveform with its annotated target span and crop the reference waveform with its reference span, then compare the paired segments. CLAP-Cos is computed on $48$\,kHz waveforms, while BEATs-Cos is computed on $16$\,kHz mono waveforms with the token axis mean-pooled into one embedding per segment. LSD is computed on the paired waveform segments from their log-power spectra, and MCD is computed with DTW-aligned MFCCs following the AC-Foley evaluation protocol. For multi-event cases, we first compute these metrics for each event, then average the event-level values within a case, and finally average over all cases in the split. Desync is computed once per generated clip against the full input video rather than per event.

\noindent \textbf{Fair comparison.} For baseline comparison, we follow the same benchmark split and input video for all methods, but adapt the reference interface of each baseline to the input format it supports. \textbf{This is particularly important for AC-Foley, whose original design assumes a single reference track and therefore cannot directly encode multiple event-specific references.} Accordingly, on each benchmark, we evaluate AC-Foley under multiple reference-construction settings to cover its default usage as well as more informative single-reference and composite-reference variants. On \EvalName{}, we report three settings: \emph{Tail}, which uses the final $2$\,s of the ground-truth clip as the default reference condition; \emph{random.}, which randomly selects one event reference; and \emph{Comp.}, which concatenates all event references in temporal order into a single composite reference. On \VividBench{}, ground-truth audio is unavailable, so the \emph{Tail} setting is not applicable. We instead report \emph{Sing.}, which uses only the reference associated with the first event, and \emph{Comp.}, which concatenates all event references in temporal order. These variants provide a broader and fairer view of how AC-Foley behaves under event-level and multi-event reference conditioning.

\subsection{VividBench Evaluation Protocol}
\label{sec:supp_vividbench_protocol}

\VividBench{} is built from VGGSound videos under five broad themes: human, animal, nature, action, and music. We first retrieve candidate videos under these themes and then manually filter them to obtain $50$ source videos with clear multi-event structure and visually identifiable sound sources. For each selected video, we manually annotate the candidate timbre themes and their corresponding temporal spans, and then retrieve an external reference waveform for each annotated event from Freesound. Based on these source videos, we evaluate separate two-event and three-event compositions. For each event, the generated audio segment is compared with its designated reference using CLAP and BEATs cosine similarity; the reported score averages the event-level values over all cases in the corresponding split. Desync is computed against the annotated video-event timing and is reported over the complete split.

\subsection{Qualitative Case Visualization}
\label{sec:supp_qual_vis}

For completeness, we provide the full qualitative visualization used in the paper as a supplementary figure. The visualization aligns sampled video frames, event-level reference spectrograms, and generated spectrograms. \textbf{You can also refer to media materials for more examples.}

\begin{figure*}[t]
    \centering
    \includegraphics[width=0.9\textwidth]{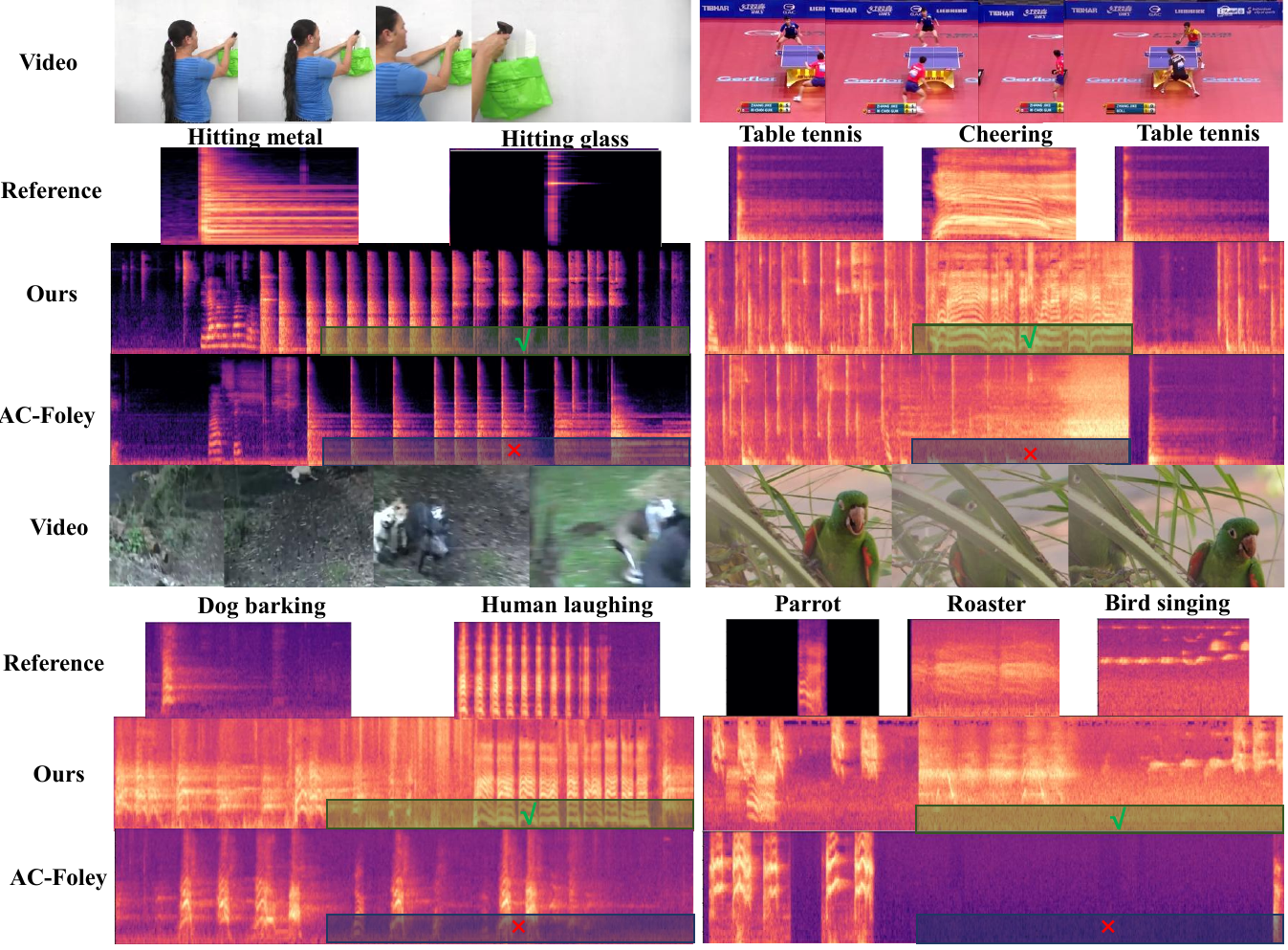}
    \caption{\textbf{Qualitative visualization on \VividBench{}.} From top to bottom, the rows show sampled video frames, event-level reference spectrograms, outputs from \ModuleName{}, and outputs from AC-Foley. The aligned layout highlights whether the generated soundtrack follows the designated event reference at the correct temporal slot and whether interference appears across neighboring events.}
    \label{fig:supp_vividbench_visualization}
\end{figure*}

\subsection{MLLM Preference Evaluation}
\label{sec:supp_user_study}

We evaluate MLLM preference on the $100$ \VividBench{} cases used in the main paper, including $50$ two-event cases and $50$ three-event cases. Each case contains the original video, event-specific reference sounds, target time spans, and anonymous candidate soundtracks generated by the five evaluated methods. Candidate identities are hidden and their order is randomized independently for each case.

We use Gemini2.5-pro as the MLLM judge and query it independently for four criteria: video--event alignment, audio quality, reference-timbre controllability, and overall preference. For each criterion, the model ranks all anonymous candidates within the same case rather than assigning absolute Likert scores. Ranking avoids calibration drift across cases and better matches the comparative nature of subjective preference evaluation. We convert a rank $r\in\{1,\ldots,5\}$ to a normalized preference score $(5-r)/4$, where higher is better, and also report the overall top-1 rate. The reported figure averages the normalized preference scores over the $100$ cases.

In each ranking call, the judge receives one time-aligned composite reference audio together with the candidate videos in randomized order. The prompt provides the case metadata, including event labels and the corresponding target spans, and asks the model to rank all candidate IDs from best to worst for alignment, quality, timbre control, and overall performance. The prompt explicitly instructs the judge to compare each target span against the reference audio placed at the same temporal interval in the composite reference, to treat timbre control as coarse perceptual reference following rather than exact recording imitation, and to avoid ties. The four ranking dimensions are queried independently so that judgments of quality, alignment, and timbre control do not interfere with one another. We then convert the returned ranking lists into per-method ranks and top-1 indicators and aggregate them over all cases.

\subsection{Long-Form V2A Application}
\label{sec:supp_longform}

We use $55$ held-out videos of duration at least $16$\,s, with at most five videos sampled from each subcategory. Each video is divided into two non-overlapping clips, $V_A=[0,8]$\,s and $V_B=[8,16]$\,s. Both systems receive the same visual inputs and synthesize one $8$\,s audio segment for each clip. MMAudio generates the two segments independently. \ModuleName{} uses the generated audio from $V_A$ as the audio context for $V_B$ through its global and condition-level reference pathways, and additionally supplies the final $2$\,s of the first segment as a tail-to-head inpainting prefix at the boundary. The two generated segments are concatenated without additional post-processing.

We quantify atmosphere consistency with LAION-CLAP cosine similarity between temporal windows of the resulting $16$\,s soundtrack. The \emph{Strict} score compares the two non-overlapping $8$\,s halves, $[0,8]$\,s and $[8,16]$\,s. The \emph{Sliding} score averages pairwise similarity over three $8$\,s windows starting at $0$, $4$, and $8$\,s, and is therefore less sensitive to a single boundary. We report the mean score for each method and the paired per-video difference $\Delta=S^{\mathrm{Ours}}-S^{\mathrm{MMAudio}}$ with its $95\%$ confidence interval. The observed improvements are $+0.085$ for Strict, with CI $[+0.035,+0.135]$, and $+0.054$ for Sliding, with CI $[+0.019,+0.088]$; both paired tests satisfy $p<0.01$.

\begin{table}[!tb]
\centering
\resizebox{\columnwidth}{!}{%
\begin{tabular}{lcc}
\toprule
Method & $S_{\text{atmos}}$ (Strict) $\uparrow$ & $S_{\text{atmos}}$ (Sliding) $\uparrow$ \\
\midrule
MMAudio                       & $0.603$ & $0.721$ \\
\textbf{\ModuleName{} (Ours)} & $\mathbf{0.688}$ & $\mathbf{0.774}$ \\
\midrule
$\Delta$ (95\% CI) & $+0.085^{*}$ $[+0.035, +0.135]$ & $+0.054^{*}$ $[+0.019, +0.088]$ \\
\bottomrule
\end{tabular}%
}
\caption{Long-form V2A with audio context on 55 held-out videos. Both models synthesize audio for two \emph{disjoint} $8$\,s video windows and concatenate the outputs into a $16$\,s soundtrack. $S_{\text{atmos}}$ is the CLAP cosine similarity between temporal windows. $\Delta$ is the paired per-video improvement with its $95\%$ confidence interval. * denotes $p<0.01$ under a paired $t$-test.}
\label{tab:app_longform}
\end{table}

\section{Training Data Annotation Summary}
\label{sec:supp_data_annotation}

This supplementary section documents the training-time annotation assets used by \ModuleName{}.
Our SFT pipeline reads three event-annotated corpora in the current codebase: VGGSound, AudioCaps, and an internal corpus.
For each source clip, the annotation file stores event boundaries, a coarse sound label, a free-form description, a confidence score, and the intra-video grouping fields used to build reference-target pairs.

\begin{table*}[t]
\centering
\caption{Training-data scale and annotation coverage. Coverage denotes the fraction of training samples that retain at least one valid event annotation.}
\label{tab:supp_train_data_summary}
\resizebox{\textwidth}{!}{%
\begin{tabular}{lrrrrrrrrrr}
\toprule
Corpus & \#Samples & \#Annotated & \#Valid Samples & \#Events & \#Valid Events & Events/Sample & Valid Events/Valid Sample & Dur. Mean & Conf. Mean & Coverage \\
\midrule
VGGSound & 181,516 & 181,516 & 149,629 & 694,819 & 694,819 & 3.83 & 4.64 & 0.76 & 0.85 & 82.4\% \\
AudioCaps & 89,269 & 89,269 & 72,387 & 288,857 & 288,857 & 3.24 & 3.99 & 0.73 & 0.83 & 81.1\% \\
Internal & 248,848 & 248,422 & 172,272 & 777,678 & 777,678 & 3.13 & 4.51 & 0.85 & 0.84 & 69.2\% \\
\bottomrule
\end{tabular}%
}
\end{table*}

\begin{table*}[t]
\centering
\caption{Event-group statistics after intra-video clustering. Similarity is measured as cosine in the LAION-CLAP-HTSAT audio embedding space (\texttt{laion\_clap}, \texttt{music\_speech\_audioset\_epoch\_15\_esc\_89.98} checkpoint), where each event embedding is obtained by feeding the audio segment inside the event span. Statistics are computed on up to 500 grouped clips per corpus. Intra: same-cluster within a video; Inter: different-cluster within a video; Cross-Video: random events sampled from different videos (5,000 pairs).}
\label{tab:supp_train_cluster_stats}
\resizebox{\textwidth}{!}{%
\begin{tabular}{lrrrrrrrrrrr}
\toprule
\multirow{2}{*}{Corpus} & \multirow{2}{*}{\#Grouped} & \multirow{2}{*}{\#Clusters} & \multirow{2}{*}{Cluster Size} & \multirow{2}{*}{Singleton} & \multirow{2}{*}{Neighbor} & \multicolumn{3}{c}{Similarity (mean$\pm$std)} & \multicolumn{2}{c}{Margin} & \multirow{2}{*}{Top-1} \\
\cmidrule(lr){7-9} \cmidrule(lr){10-11}
  &   &   &   &   &   & Intra & Inter & Cross-Vid & Intra$-$Inter & Intra$-$Cross &   \\
\midrule
VGGSound & 112,398 & 307,211 & 1.67 & 51.1\% & 1.12 & 0.773$\pm$0.132 & 0.702$\pm$0.157 & 0.312$\pm$0.138 & 0.071 & 0.461 & 0.785 \\
AudioCaps & 50,963 & 126,418 & 1.66 & 49.7\% & 1.08 & 0.751$\pm$0.135 & 0.659$\pm$0.159 & 0.338$\pm$0.146 & 0.092 & 0.412 & 0.756 \\
Internal & 128,457 & 340,739 & 1.66 & 50.9\% & 1.10 & 0.808$\pm$0.118 & 0.753$\pm$0.134 & 0.396$\pm$0.149 & 0.055 & 0.413 & 0.812 \\
\bottomrule
\end{tabular}%
}
\end{table*}

Table~\ref{tab:supp_train_data_summary} shows that all three corpora provide large-scale event-level supervision, but with different trade-offs between cleanliness and diversity. VGGSound and AudioCaps retain more than $80\%$ valid samples, indicating relatively stable event annotations after filtering. In contrast, the internal corpus has lower coverage, reflecting more complex acoustic composition, but it contributes the largest number of samples and events overall. This makes it an important source of multi-event diversity even though its raw annotations are noisier. We also observe that the valid-event density remains high across corpora, which means that each retained sample typically still contains multiple usable events for constructing reference-target training pairs rather than collapsing into near-single-event supervision.

Table~\ref{tab:supp_train_cluster_stats} further validates the intra-video grouping used by our data construction pipeline. Across all corpora, same-cluster pairs consistently have higher similarity than both different-cluster pairs from the same video and random cross-video pairs, showing that the grouping captures meaningful acoustic affinity instead of arbitrary temporal co-occurrence. At the same time, the margin between Intra and Inter is noticeably smaller than the margin between Intra and Cross-Video, especially for the internal corpus. This suggests that events from the same video still share substantial background or recording characteristics even when they belong to different clusters, which is exactly the challenging regime where event-level timbre control is needed. Finally, the singleton ratio stays around $50\%$ and the average neighbor count is close to one, indicating that direct same-group replacement is useful but limited; this motivates the additional speed, mix, degradation, and empty-reference perturbations used to diversify the reference side during training.

\subsection{Annotation Schema}

Each event file follows the same JSON structure across corpora:

\begin{verbatim}
{
  "video": "...mp4",
  "audio": "...wav",
  "num_events": 3,
  "events": [
    {
      "start": 1.18,
      "end": 1.98,
      "energy": 0.63,
      "label": "music",
      "desc": "soft ambient melody",
      "confidence": 0.80,
      "group": 0,
      "neighbors": [3, 4]
    }
  ]
}
\end{verbatim}

The main fields are used as follows.
\texttt{start}/\texttt{end} specify the temporal span of the event in the source clip;
\texttt{energy} stores the segment energy used by the detector;
\texttt{label}, \texttt{desc}, and \texttt{confidence} come from the multimodal annotator;
\texttt{group} is the intra-video cluster identifier; and \texttt{neighbors} lists group-associated events that can be reused as timbre-consistent references.

\subsection{From Event Annotations to Training Samples}

The event annotations above are not fed to the generator directly. Instead, we convert them into event-conditioned training records that explicitly distinguish reference-side timing from target-side timing. For each valid event, the target span is always the event's own annotated interval, while the reference side is constructed from the same perturbation family described in the main paper: same-group replacement, speed variation, mixing, acoustic degradation, together with an optional empty-reference condition. Same-group replacement uses one item from \texttt{neighbors}; if no valid neighbor exists, the implementation falls back to another valid construction path.

The empty mode deserves special attention. When an event is sampled as empty, we do not insert a placeholder embedding or a special token. Instead, the corresponding reference slot is left silent when constructing TACO-Ref. As a result, one event in a multi-event sample may receive a valid acoustic reference while another event in the same sample receives an explicit zero-reference condition. This teaches the model that reference support can be partial and prevents it from overfitting to the assumption that every target event must always be paired with an informative exemplar.

A typical exported training row contains fields such as
\texttt{rag\_audio\_list}, \texttt{rag\_seg\_in\_ref}, and \texttt{rag\_seg\_in\_gen}, which describe which waveform is used as the event reference, which segment is read from that waveform, and where the resulting reference should be expressed in the generated soundtrack. When inpainting or background anchoring is enabled, the row additionally stores \texttt{inp\_audio\_list}, \texttt{inp\_seg\_in\_ref}, and \texttt{inp\_seg\_in\_gen} for the unmarked regions. This separation between reference-side and generation-side spans is what allows us to build a time-aligned composite reference while keeping the training data fully reproducible.

\end{document}